\documentclass[prd,
 notitlepage,
nofootinbib,
amsfonts,
amssymb,
amsmath,
onecolumn,
 ]{revtex4-2}
\usepackage{graphicx}
\usepackage{color}  
\usepackage[normalem]{ulem}
\usepackage[dvipsnames]{xcolor}
\usepackage{scalerel,stackengine}
\usepackage[T1]{fontenc}
\usepackage{soul}

\definecolor{bluc}{cmyk}{1,0.9,0,0}
\definecolor{rossoCP3}{cmyk}{0,.88,.77,.40}
\definecolor{rosso}{cmyk}{0,1,1,0.4}
\definecolor{rossos}{cmyk}{0,1,1,0.55}
\definecolor{rossoc}{cmyk}{0,1,1,0.2}
\definecolor{verdes}{cmyk}{0.92,0,0.59,0.4}

\usepackage{hyperref}
\hypersetup{colorlinks, bookmarksopen, bookmarksnumbered, citecolor=verdes, linkcolor=bluc, pdfstartview=FitH, urlcolor=rossos}

\def\ga{\mathrel{\raise.3ex\hbox{$>$\kern-.75em\lower1ex\hbox{$\sim$}}}}
\def\la{\mathrel{\raise.3ex\hbox{$<$\kern-.75em\lower1ex\hbox{$\sim$}}}}
\def\beq{\begin{equation}}
\def\eeq{\end{equation}}
\def\be{\begin{equation}} 
\def\ee{\end{equation}}
\def\bea{\begin{eqnarray}}
\def\eea{\end{eqnarray}}

\def\nn{\nonumber}

\begin{document}

\title{On the entropy of a stealth vector-tensor black hole}
\author{Javier Chagoya}
\author{I. D\'iaz-Salda\~na}
\author{J.C. L\'opez-Dom\'inguez}
\author{C. Mart\'inez-Robles}

\affiliation{Unidad Acad\'emica de F\'isica, Universidad Aut\'onoma de Zacatecas, \\ Calzada Solidaridad esquina con Paseo a la Bufa S/N 
C.P. 98060, Zacatecas, M\'exico.}

\begin{abstract}
We apply Wald's formalism to a Lagrangian within generalised Proca gravity that admits a Schwarzschild black hole with a non-trivial vector field. The resulting entropy differs from that of 
the same black hole in General Relativity by a logarithmic correction modulated 
by the only independent charge of the vector field. We find conditions on this charge to guarantee that the entropy is a non-decreasing function of the black hole area, as is the case in GR. If this 
requirement is extended to 
black hole mergers, we find 
that for Planck scale black holes, a non-decreasing
entropy is possible only if the area of the final black hole is several times larger than the initial total area of the merger. Finally, we discuss some implications of the vector Galileon entropy from the point of view of entropic gravity.
\end{abstract}

\date{\today}
\maketitle
\tableofcontents
\section{Introduction}
Since its conception, the theory of General Relativity (GR) proposed by Albert Einstein in 1917, has been subject
to a variety of observational tests. In the
realm of cosmology, the concordance model is $\Lambda$CDM,
which is based on GR. This model is in agreement with
most cosmological observations that constrain
the evolution and properties of the background spacetime, the matter fields and their perturbations~\cite{Planck:2015fie}.
In addition, recent astrophysical observations
show compatibility with the properties of the spacetime around black holes predicted by GR~\cite{EventHorizonTelescope:2019dse,LIGOScientific:2019fpa, LIGOScientific:2020tif, EventHorizonTelescope:2022wkp}. {In this
scenario, alternative theories of gravity need to 
predict cosmic and astrophysical solutions that are 
similar to the spacetimes predicted by GR in order to be considered viable.} 
Several theories of modified gravity predict
solutions that deviate parametrically
from GR, and these deviations can be used to put
constraints on the parameter(s) of the theory by 
contrasting their predictions to common cosmological and
astrophysical direct or indirect observables, such as luminosity distance, Hubble parameter evolution, structure
formation, deflection of light and angular diameter of
black hole shadows. However, some other theories
of modified gravity predict solutions where the metric 
coincides exactly with GR solutions even though there is a non-trivial profile for the additional fields of the theory. These are known in the literature as \textit{stealth} solutions~(e.g.~\cite{Babichev:2013cya,Kobayashi:2014eva,Cisterna:2016nwq,Minamitsuji:2016ydr,Chagoya:2016aar,Babichev:2017guv,BenAchour:2018dap,Minamitsuji:2019shy,Motohashi:2019sen,Takahashi:2020hso}). 
Models with this type of solutions can be more difficult to rule out as alternative theories of gravity, for instance, they can evade solar system constraints. Amongst the possibilities to discriminate these theories, one could consider a direct coupling between the additional fields and matter, or study perturbations. Indeed, at the perturbative level, couplings between the extra fields and matter can appear automatically, as shown in~\cite{deRham:2019gha} for a scalar-tensor model. Another possibility recently studied in the context of degenerate higher order scalar-tensor theories of gravity, is to introduce a small detuning of the degeneracy conditions, this is known as the {\it scordatura} mechanism (e.g.~\cite{DeFelice:2022qaz}). In this work we explore yet another possibility, namely, 
we analyse the Wald entropy of the black hole, which we expect to differ from the entropy of a GR Schwarzschild black hole since, even though the metric is the same, the Lagrangian is 
not the same as in GR and the vector field is not trivial. 
In particular, we work with a model
that belongs to generalised Proca theories, also known as vector Galileons, a type of
theories that might be relevant for explaining the
accelerated expansion of the universe~\cite{Heisenberg:2014rta, Tasinato:2014eka}, and have 
been shown to predict a rich phenomenology for
compact objects~\cite{Chagoya:2016aar,Fan:2016jnz,Minamitsuji:2016ydr,Cisterna:2016nwq,Chagoya:2017fyl,Babichev:2017rti,Heisenberg:2017xda,Heisenberg:2017hwb,Fan:2017bka}. We study the Wald entropy of this black hole, which turns out to be the Bekenstein-Hawking entropy with a logarithmic correction 
introduced by the vector field. We analyse the consequences of this correction in two different scenarios. The first one is related to the area theorem of GR~\cite{Hawking:1971vc}, and motivated by recent observational support to this theorem~\cite{PhysRevLett.127.011103} using gravitational waves. The second scenario we analyse is in the context of the entropic gravity derivation of a modified Newtonian potential~\cite{Verlinde:2010hp, Modesto:2010rm} and a modified Friedmann equation~\cite{Cai:2008ys,Sheykhi:2010yq,Sheykhi:2010zz}.

This work is organised as follows. In Sec.~\ref{sec:bhvg}
we present the vector Galileon model and black hole 
solution under study. In Sec.~\ref{sec:wald} we review the
basics of Wald formalism and its application to the
Schwarzschild black hole in GR, and then we apply the 
formalism to the Schwarzschild black hole in VG
and discuss some properties of the resulting entropy. In Sec.~\ref{sec:entro} we explore the 
implications of the VG modified entropy for the area theorem and for astrophysical and cosmological applications of entropic gravity. Finally, Sec.~\ref{sec:dis} is devoted to conclusions.


\section{Black holes in vector-tensor gravity}\label{sec:bhvg}
We focus on the theory of vector-tensor gravity known as \textit{generalized Proca theory}, also referred to as vector galileons due to the similarity between their construction and that of scalar galileons. As in Proca theory, the longitudinal mode of the vector field becomes dynamical as a consequence of breaking the Abelian symmetry. However, in generalised Proca the Abelian symmetry is broken not only by the mass term of the vector field but also by {non--minimal} couplings with gravity. This characteristic offers a way around Bekenstein's no-go theorem, which applies to black holes in theories where the Abelian symmetry is broken by a mass term. This was 
exploited in~\cite{Chagoya:2016aar} to show that the Schwarzschild spacetime is a solution in a sector of generalised Proca theory. More general black hole solutions have been reported in~\cite{Fan:2016jnz,Minamitsuji:2016ydr,Cisterna:2016nwq,Chagoya:2017fyl,Babichev:2017rti,Heisenberg:2017xda,Heisenberg:2017hwb,Fan:2017bka}. In this work, we use the stealth Schwarzschild solution that is obtained in the model described by
\begin{equation}
 S
 = \int d^4x \sqrt{-g}\left[{\frac{1}{16\pi}}\, R  -\frac{1}{4} {F_{ab}F^{ab}} + \gamma\,G_{ab}A^aA^b \right],\label{actG}
\end{equation}
where the electromagnetic tensor $F_{ab}=\partial_a A_b - \partial_b A_a$ is constructed 
out of the 4--potential $A_b$ and $R$ is the Ricci scalar of the metric $g_{ab}$. 
The coupling {of the 4--potential $A_b$} with the Einstein tensor $G_{ab}$, weighted by the dimensionless constant $\gamma$,   
breaks the Abelian symmetry of the Lagrangian and turns on the longitudinal mode of the vector field 
in such a way that the theory propagates five degrees of freedom~\cite{Heisenberg:2014rta}.
Considering a spherically symmetric ansatz,
\begin{eqnarray}
ds^2 &=& -f(r) dt^2 + h(r)^{-1} dr^2 + r^2 d\theta^2 + r^2 \sin^2\theta d \varphi^2,  \nonumber\\
A_b &=& (A_0(r), A_1(r), 0,0),
\label{eq:metric}
\end{eqnarray}
the equations of motion for the system can be expressed as 
\begin{subequations}\label{eqs}
\begin{eqnarray}
0&=& \gamma A_1\left[h-f (f+rf')^{-1}\right] , \label{eqs1} \\
0&=&  {\frac{ f}{4\pi}}( h-1) + h r^2 {A_0'}^2+ 2 {\frac{1}{8\pi}} r h  f' \nn\\ 
&&+\gamma  \left[2 A_0^2 (h-1)+2 f h (3 h-1) A_1^2+8 {A_0} h r {A_0}'
-2 A_0^2 h r \frac{f'}{f}+6 h^2 A_1^2 r f'\right]
, \label{eqs2} \\
0&=&{\frac{ f}{4\pi}}(1-h-rh') - h r^2 {A_0'}^2 
\nn \\ 
&&+\gamma  \left[
2 A_0^2 \left(h-1+r h'\right)-2 f h {A_1} \left({A_1} \left(h+1+3 r h'\right)+4 h r {A_1'}\right)\right] ,
\label{eqs3} \\
0&=&
4 {A_0} \gamma  \left(h-1+r h'\right)
-r \left[{A_0'} \left(h r \frac{f'}{f}-4 h-r h'\right)-2 h r {A_0''}\right]
,
\label{eqs4}
\end{eqnarray}
\end{subequations}
where a prime indicates derivative {with respect to the radial coordinate}. The radial component
of the vector field, $A_1(r)$,  is associated with the longitudinal polarization of the vector, which 
has an important role in our scenario that breaks the Abelian symmetry. If $\gamma=0$ or $A_1(r)=0$,
Eq. \eqref{eqs1} is identically satisfied, and the remaining equations lead to the Reissner-N\"ordstrom solution. On the other hand, if $\gamma\neq0$ and $A_1(r)\neq0$, 
the first equation fully determines $h(r)$ in terms of $f(r)$. Asymptotically flat solutions for
arbitrary values of $\gamma$ can be found analytically and numerically. However, only for 
$\gamma = 1/4$ Minkowski spacetime is an exact solution with a non-trivial profile of
the vector field. This observation motivates {us to set}  $A_1(r)\neq 0$
and $\gamma = 1/4$. Under these assumptions, 
Eqs.~(\ref{eqs}) lead to the following solutions:
\begin{align}
f(r)&=h(r)\,=\,1-\frac{2 M}{r} , \label{g1f}
\\
A_0(r)&=\frac{Q}{r}+P ,\label{g1a0}
\\
A_1(r)&=\frac{\sqrt{Q^2+2\,r\,P(Q+M\,P)}}{ r-2 M }  ,\label{g1pi}
\end{align}
where $M$ represents the black hole mass, and $Q,~P$ are charges associated to the vector field. In particular, $P$ determines
the asymptotic behaviour of $A_1(r)$: if $P=0$ then $A_1(r) \sim 1/r$, while if $P\neq0$ then $A_1(r) \sim 1/\sqrt{r}$. Furthermore, $P$
also determines the norm of the vector field, which for this solution is a constant, $A_b A^b = -P^2$~\footnote{A systematic study of black hole solutions to generalised Proca theories under the condition $A_b A^b = constant$ is presented in~\cite{Heisenberg:2017xda}.}. Notice that
the metric components, Eq.\eqref{g1f}, exactly coincide with the Schwarzschild space-time, thus this is a {\it stealth} solution, i.e. it matches a GR counterpart, but with a non-trivial profile for the vector field. In the next sections we explore how the entropy of this black hole 
{mismatches} that of a Schwarzschild black hole in General Relativity. 
\section{Wald formalism}\label{sec:wald}

In this section we review the notion of black hole entropy as an integral of a N\"oether charge, introduced by R. Wald in 1993~\cite{Wald:1993nt,Iyer:1994ys}. Let us consider an action $I$ given as the integral of a Lagrangian $4$-form $\mathbf L$,
\begin{equation}
I = \int \mathbf L = \int \boldsymbol{\epsilon} L,
\end{equation}
where $\boldsymbol\epsilon$ is the canonical volume form on the 4-dimensional manifold.
Assuming that the Lagrangian is diffeomorphism covariant, it can be expressed as 
\begin{equation}
    \mathbf L = \mathbf L\left(
    g_{ab}, R_{bcde} , \nabla_{a_1} R_{bcde}, \dots,
    \nabla_{(a_1\dots }\nabla_{a_m)} R_{bcde}, \Psi,  \nabla_{a_1} \Psi, \dots,
    \nabla_{(a_1\dots }\nabla_{a_m)} \Psi
    \right),
\end{equation}
where {the parentheses around indices indicate symmetrization} 
and 
$\Psi$ represents any dynamical fields with arbitrary index 
structure, for instance a scalar or a vector field. The first
order variation of the Lagrangian can be expressed, after
integration by parts, as
\begin{equation}
\delta \mathbf L = (\mathbf E_g)^{ab}\delta g_{ab} + \mathbf E_\Psi \delta\Psi + d\mathbf\Theta,
\end{equation}
where all the variations $\delta\Psi$ of the additional fields are summed over and the boundary term is given by the 3-form $\mathbf\Theta$ that depends on the fields, their variations, and their derivatives. The dependency of $\mathbf\Theta$ on the variations is only linear. From the requirement that the variation of the action vanishes, the equations of motion are
\begin{equation}
(\mathbf E_g)^{ab} = 0, \ \ \ \mathbf E_\Psi = 0
\end{equation}
The boundary terms coming from the dependencies of the Lagrangian on
the metric and Riemann tensors can be worked out in general~\cite{Iyer:1994ys}, so that 
the 3-form $\mathbf\Theta$ can be written as
\begin{equation}
  \mathbf\Theta = 2 \mathbf{E}_R^{bcd}\nabla_d\delta g_{bc}
  - 2 (\nabla_d\mathbf{E}_R^{bcd})\delta g_{bc}
  +\tilde{\mathbf\Theta},\label{eq:theta}
\end{equation}
where $\tilde{\mathbf\Theta}$ contains the boundary terms
associated to the additional fields $\Psi$ and, restricting our attention to Lagrangians that do not depend on derivatives of the Riemann tensor, {we have}
\begin{equation}
    (\mathbf{E}_R^{bcd})_{a_1 a_2 a_3} = \epsilon_{a a_1 a_2 a_3}
    \frac{\partial L}{\partial R_{abcd}},
\end{equation}
 where $\epsilon_{abcd}$ is the Levi--Civita tensor density. Under this simplifying assumption, the components of $\mathbf\Theta$ can be expressed 
as
\begin{equation}
(\mathbf\Theta)_{a_1 a_2 a_3} =  \epsilon_{a a_1 a_2 a_3}\left[ 2 \frac{\partial L}{\partial R_{abcd} }\nabla_d \delta g_{bc} - 2 \nabla_d \left( \frac{\partial L}{\partial R_{abcd} } \right) \delta g_{bc} +\tilde\Theta^a\right] \equiv  \epsilon_{a a_1 a_2 a_3} \mathcal J^a.\label{eq:3formtheta}
\end{equation}
For a diffeomorphism invariant theory, considering
variations of the fields generated by an infinitesimal vector field $\boldsymbol\xi$, there is a 
$3$-form $\mathbf J$ that defines the conserved N\"oether
current as
\begin{equation}
\mathbf J = \mathbf\Theta(g,\Psi,\mathcal L_\xi g,\mathcal L_\xi \Psi) - \boldsymbol\xi\cdot\mathbf L, 
\label{eq:j3form}
\end{equation}
where $\mathbf\Theta(g,\Psi,\mathcal L_\xi g,\mathcal L_\xi \Psi)$  is obtained {from Eq.~(\ref{eq:3formtheta})} considering that the variations of the fields are given by their Lie derivatives along the infinitesimal generator of the
diffeomorphism, and $\boldsymbol\xi\cdot\mathbf L$ represents the
contraction of the vector field components $\xi^a$ with the first index
of the Lagrangian 4-form $\mathbf L$. It can be shown that
$\mathbf J$ is a closed form \textit{under the condition that
the equations of motion are satisfied}. In this case, it can be expressed as
\begin{equation}
    \mathbf J = d \mathbf Q.\label{eq:N\"oethercharge}
\end{equation}
The 2-form $\mathbf Q$ is called the 
{\it N\"oether charge}. As shown in~\cite{Wald:1993nt}, when the linearized 
field equations are satisfied, the variation of the
Hamiltonian associated to variations of the fields generated by $\xi^a$
is given by
\begin{equation}
    \delta\mathcal H =\int_\infty (\delta \mathbf Q - \boldsymbol{\xi}\cdot\boldsymbol\Theta)
    \label{eq:deltah},
\end{equation}
where integration is performed over a 2-dimensional sphere at infinity. 
Let us consider static black holes with a bifurcate killing horizon $\Sigma$, i.e., 
a surface where the killing field vanishes. For the Schwarzschild space-time
this surface coincides with the position of the event horizon. In this setup,
the first law of thermodynamics derived in~\cite{Wald:1993nt} (see also~\cite{Feng:2015oea}) can be
expressed as
\begin{equation}
    \int_\infty (\delta \mathbf Q - \boldsymbol{\xi}\cdot\boldsymbol\Theta) =  \int_\Sigma (\delta \mathbf Q - \boldsymbol{\xi}\cdot\boldsymbol\Theta).\label{eq:1stlaw}
\end{equation}
Choosing $\boldsymbol\xi$ as the timelike killing vector, Eq.~\eqref{eq:deltah}
gives the change $\delta M$ in the ADM mass of the solution. On the other hand, the right hand side of 
Eq.~\eqref{eq:1stlaw} is defined as $(\kappa/2\pi)\delta S$, where $S$ is the entropy and
$\kappa$ the surface gravity of the solution, related to the temperature by
$T =\kappa/2\pi $. Thus, Eq.~\eqref{eq:1stlaw} can be expressed as $\delta M = T \delta S$. From here, one can obtain the black hole entropy.

In order to exemplify the application of Wald formalism, let us review the computation of the entropy for a Schwarzschild black hole in GR. In geometrized units, the components of the Lagrangian 4-form are 
\begin{equation}
    (\mathbf L)_{abcd} =
    \frac{1}{16\pi} \epsilon_{abcd} R ={ \epsilon_{abcd} 
L}.
\end{equation}
Notice that the factor $1/16\pi$ is included in $L$. Using the symmetry properties of the metric
and Riemann tensor, it is straightforward to verify that 
\begin{equation}
    \frac{\partial L}{\partial R_{abcd}} = \frac{1}{32\pi} \left( g^{ac} g^{bd} - g^{ad} g^{bc} \right).
\end{equation}
Therefore, the $\mathcal J^a$ defined in Eq.~(\ref{eq:3formtheta}) reduces to
\begin{equation}
    \mathcal J^a = {\frac{1}{16\pi}}\left( g^{ac} g^{bd} - g^{ad} g^{bc} \right)\nabla_d \delta g_{bc} ={\frac{1}{16\pi}} g^{ad}g^{bc} \left(
    \nabla_c \delta g_{bd} - \nabla_d \delta g_{bc}
    \right).\label{eq:j1form}
\end{equation}
To construct the 3-form $\boldsymbol\Theta$ we need to consider variations of the
metric generated by a diffeomorphism, i.e., 
\begin{equation}
    \delta g_{bc} =  \nabla_b \xi_c + \nabla_c \xi_b.\label{eq:metdiff}
    \end{equation}
Inserting Eqs.~(\ref{eq:j1form},\ref{eq:metdiff})
in the equation for $\mathbf J$, Eq.~(\ref{eq:j3form}), we get
\begin{align}
    (\mathbf J)_{ijk} & =
    \frac{1}{8\pi}  \epsilon_{ijka}\left[\nabla_b  \nabla^{[a} \xi^{b]}   - \xi_b  
    G^{ab}\right],
\end{align}
where the square brackets around the indices represent antisymmetrization and $G^{ab}$ is the Einstein tensor.
When the equations of motion of GR in vacuum are satisfied, this tensor vanishes. 
Therefore, the 2-form N\"oether charge $\mathbf Q$,
defined in Eq.~(\ref{eq:N\"oethercharge}) comes from the first term in the previous equation. One can verify that its components
\begin{equation}
   ( \mathbf Q)_{ij} = -{\frac{1}{16\pi}}\epsilon_{ijab} \nabla^a\xi^b,
\end{equation}
are such that
\begin{align}
    (d \mathbf Q)_{ijk} 
    = - {\frac{1}{16\pi}} 3 \nabla_{[i} \left(\epsilon_{jk]ab}\nabla^a\xi^b \right) = \mathbf J_{ijk},
\end{align}
when the equations of motion are satisfied. Now we evaluate $\delta \mathbf Q$ for a metric of the form given in Eq.~(\ref{eq:metric}). It is important to highlight that, while $\mathbf Q$ is constructed for a diffeomorphism, its variations in Eq.~(\ref{eq:1stlaw})
are only subject to the condition that they satisfy the linearized equations of motion. 
We obtain these variations by
replacing $f(r) \to f(r) + \delta f(r)$ 
and 
$h(r) \to h(r) + \delta h(r)$ 
in Eq.~(\ref{eq:metric}), where $\delta h$ and $\delta f$ are to be specified depending on whether we are evaluating these
quantities at infinity or at the Killing horizon $\Sigma$. 
Thus, we find that the non-vanishing components of $\delta\mathbf Q - \boldsymbol\xi\cdot\boldsymbol\Theta$ are 
\begin{equation}
(\delta \mathbf Q -\boldsymbol\xi\cdot\boldsymbol\Theta)_{23}  = -(\delta\mathbf Q -\boldsymbol\xi\cdot\boldsymbol\Theta)_{32} = -{\frac{r}{8\pi}}{ }\sqrt{\frac{f(r)}{h(r)}}\delta h(r).
\end{equation}
Specializing to a Schwarzschild metric, the variations of $h(r)$ at infinity and at the horizon are, respectively,
\begin{align}
\delta h(r_\infty) & = - \frac{2 \delta m}{r},  \\
\delta h(r_h) & = - \frac{2 m \delta r_h}{r_h^2}.
\end{align}
Thus, the right hand side of Eq.~(\ref{eq:1stlaw}) leads to 
\begin{equation}
\frac{1}{2}\delta r_h = \frac{\kappa}{2\pi} \delta S.
\label{eq:firstlaw2}
\end{equation}
For a Schwarzschild black hole the surface gravity is
$\kappa = (4m)^{-1} $. Substituting this into Eq.~(\ref{eq:firstlaw2}) and rearranging terms, we get
\begin{equation}
\delta S = \delta \left( \frac{A}{4}  \right),
\end{equation}
where $A$ is the area of a sphere with radius equal to the black hole horizon.

\subsection{Wald entropy for a VG Schwarzschild black hole }
Let us now apply the Wald formalism described and exemplified in the previous section to the vector Galileon Lagrangian introduced in Eq.~(\ref{actG}) and reproduced here for clarity,
\begin{equation}\label{th}
    L=\frac{1}{16\pi} R-\frac{1}{4}F^{ab}F_{ab} +\gamma G_{ab}A^{a}A^{b}.
\end{equation}
As discussed in Sec.~\ref{sec:bhvg}, this theory admits a Schwarzschild solution with a non-trivial vector field, given in Eqs.~(\ref{g1f},\ref{g1a0},\ref{g1pi}). Since there is an additional field, the term $\tilde{\mathbf\Theta}$ in Eq.~(\ref{eq:theta}) becomes relevant. Following Eq.~(\ref{eq:3formtheta}), this leads
to a vector field $\mathcal J^a$
given by 
\begin{equation}
\mathcal{J}^a = 2 \frac{\partial L}{\partial R_{abcd} }\nabla_d \delta g_{bc} - 2 \nabla_d \left( \frac{\partial L}{\partial R_{abcd} } \right) \delta g_{bc} 
+\frac{\partial L}{\partial(\nabla_{a}A_{b})}\delta A_{b}.
\end{equation}
Notice that the vector Galileon part of the Lagrangian does not 
contribute to the third term of $\mathcal J^a$, but it does
contribute to the first two terms through the Riemann tensors contained in $G_{ab}$.
As reviewed in the previous section, for the GR Lagrangian one gets
\begin{equation}
    \mathcal J_{GR}^a  = {\frac{1}{16\pi}}g^{ad}g^{bc} \left(
    \nabla_c \delta g_{bd} - \nabla_d \delta g_{bc}
    \right),\label{eq:j1formgr}
\end{equation}
while for the Maxwell and vector Galileon Lagrangian, namely  
$L_{A}=-\frac{1}{4}F_{ab}F^{ab}+\gamma G_{ab}A^{a}A^{b}$, we have
\begin{align} \mathcal J^{{a}}_{A}=-F^{{a}{b}}\delta A_{{b}}+\gamma \biggl\{ &-8\pi A^2  \mathcal J_{GR}^{a}+\frac{1}{2} g^{{a} {d}} g^{{b} {c}}\left[\nabla_{{c}}A^{2} \delta g_{{b} {d}}-\nabla_{{d}}A^{2} \delta g_{{b} {c}}\right]+\frac{1}{2} g^{{a} {b}} A^{{d}}A^{{c}} \nabla_{{d}} \delta g_{{c} {b}}-\frac{1}{2} \nabla_{{d}}\left(A^{{a}} A^{{c}} \right) g^{{d} {b}} \delta g_{{c} {b}} \nonumber \\ &-\frac{1}{2} g^{{a} {b}} A^{{d}} A^{{c}} \nabla_{{b}} \delta g_{{d} {c}}+\frac{1}{2} \nabla_{{b}}\left(A^{{d}} A^{{c}}\right) g^{{b} {a}} \delta g_{{d} {c}} -\frac{1}{2} g^{{d} {b}} A^{{a}} A^{{c}} \nabla_{{c}} \delta g_{{d} {b}}+\frac{1}{2} \nabla_{{c}}\left(A^{{c}} A^{{a}}\right) g^{{d} {b}} \delta g_{{d} {b}}\nonumber \\ &+\frac{1}{2} g^{{d} {b}} A^{{a}} A^{{c}} \nabla_{{b}} \delta g_{{d} {c}}-\frac{1}{2} \nabla_{{d}}\left(A^{{d}} A^{{c}}\right) g^{{a}{b}} \delta g_{{c} {b}} \biggl\}.
\end{align}
Then, the components of the 3-form
$\boldsymbol\Theta$ are given by
\begin{equation}\label{ThetaFull}
   ( \boldsymbol\Theta)_{a_1 a_2 a_3}=\epsilon_{a a_1 a_2 a_3}( \mathcal J^{a}_{GR}+ \mathcal J^{a}_{A} ).
\end{equation}
When the variations are generated by an infinitesimal diffeomorphism the metric transforms according 
to Eq.~(\ref{eq:metdiff}), while the vector field transforms as
\begin{equation}
    \delta A_{b}=\xi^{d}\nabla_{d}  A_{b}+A_{d}\nabla_{b}\xi^{d}.
\end{equation}
Inserting this into Eq.~(\ref{eq:j3form}) we obtain the N\"oether current $3$-form $\mathbf{J} = \mathbf{J}_{GR} + \mathbf{J}_{A}$, where
$\mathbf{J}_{GR}$ comes
from the Einstein-Hilbert Lagrangian $L_{EH} = R/16\pi$ and its components are given by
\begin{equation}
\left(\mathbf{J}_{GR}\right)_{ijk}=-{\frac{1}{8\pi}} \epsilon_{  ijk a}G^{ab}\xi_{b}+{{\frac{1}{8\pi}}} \epsilon_{ ijka}\nabla_{b}\left(\nabla^{[a}\xi^{b]}\right),
\end{equation}
and $\mathbf{J}_{A}$ comes from $L_A$
and has the components
\begin{equation}
\begin{aligned}   \left(\mathbf{J}_{A}\right)_{ijk}=2\epsilon_{ijk a}\nabla_{b} \biggl\{& \frac{1}{2}F^{ab}A^{d}\xi_{d} -\frac{1}{2}\gamma A^{2}\nabla^{[a}\xi^{b]}+\gamma \nabla^{[a}A^{2}\xi^{b]}+\frac{1}{2}\gamma A^{{d}}A^{[a}\nabla_{{d}}\xi^{b]}-\gamma \nabla_{{d}}\left(A^{{d}}A^{[a}\right)\xi^{b]}\\
  &  -\gamma \nabla^{[a}\left(A^{b]}A^{{d}}\right)\xi_{{d}}-\frac{1}{2}\gamma A^{{d}}A^{[a}\nabla^{b]}\xi_{{d}}\biggl\}+\epsilon_{ijka}T^{ab}\xi_{b}+\tilde{\mathbf{E}}_{A},
    \end{aligned}
\end{equation}
where $\tilde{\mathbf{E}}_{A}$ vanishes when the equation of motion for $A^{b}$ is satisfied, which we assume to be true, and $T^{ab}$ is the energy-momentum tensor resulting from the vector Lagrangian $L_A$, i.e.,
\begin{equation}
T^{ab}=-2\left(\frac{\partial L_{A}}{\partial g_{ab}}-\frac{1}{2}g^{ab}L_{A}\right).
\end{equation}
Collecting all the terms together we have
\begin{equation}\label{j3}
\begin{aligned}
(\mathbf{J})_{ijk}=&-{\frac{1}{8\pi}}  \epsilon_{ijk a}  
    {\left(G^{ab}-{{8\pi}} T^{ab}\right)} \xi_{b}\\
    &+2\epsilon_{ijk a}\nabla_{b}\biggl\{ {\frac{1}{16\pi}} \nabla^{[a}\xi^{b]}+\frac{1}{2}F^{ab}A^{d}\xi_{d}-\frac{1}{2}\gamma A^{2}\nabla^{[a}\xi^{b]}+\gamma\nabla^{[a}A^{2}\xi^{b]}\\
    &+\frac{1}{2}\gamma A^{d}A^{[a}\nabla_{d}\xi^{b]}-\gamma\nabla_{d}\left(A^{d}A^{[a}\right)\xi^{b]} -\gamma \nabla^{[a}\left(A^{b]}A^{d}\right)\xi_{d}-\frac{1}{2}\gamma A^{d}A^{[a}\nabla^{b]}\xi_{d}\biggl\}.
    \end{aligned}
\end{equation}
The first term in Eq.~(\ref{j3}) vanishes due to the Einstein field equations
\begin{equation}
{G^{ab}}=8\pi T^{ab},
\end{equation}
therefore, the final form of the N\"oether current is
\begin{equation}\label{j4}
\begin{aligned}
    (\mathbf{J})_{ijk}= &2\epsilon_{ijk a}\nabla_{b}\biggl\{ {\frac{1}{16\pi}} \nabla^{[a}\xi^{b]}+\frac{1}{2}F^{ab}A^{d}\xi_{d}-\frac{1}{2}\gamma A^{2}\nabla^{[a}\xi^{b]}+\gamma\nabla^{[a}A^{2}\xi^{b]}\\
    &+\frac{1}{2}\gamma A^{d}A^{[a}\nabla_{d}\xi^{b]}-\gamma\nabla_{d}\left(A^{d}A^{[a}\right)\xi^{b]} -\gamma \nabla^{[a}\left(A^{b]}A^{d}\right)\xi_{d}-\frac{1}{2}\gamma A^{d}A^{[a}\nabla^{b]}\xi_{d}\biggl\}.
    \end{aligned}
\end{equation}
As can be seen from the last expression, the $3-$form $\mathbf{J}$ is the exterior derivative of a $2$-form $\mathbf{Q}$, i.e. $\mathbf{J}=d \mathbf{Q}$, where
\begin{equation}
    \begin{aligned}\label{2-form}
    (\mathbf{Q})_{ij}=-\epsilon_{ijab}\biggl[&{\frac{1}{16\pi}} \nabla^{a}\xi^{b}+\frac{1}{2}F^{ab}A^{d}\xi_{d}-\frac{1}{2}\gamma A^{2}\nabla^{a}\xi^{b}+\gamma \nabla^{a}A^{2}\xi^{b}+\frac{1}{2}\gamma A^{d}A^{a}\nabla_{d}\xi^{b}\\
    & -\gamma\nabla_{d}\left(A^{d}A^{a}\right)\xi^{b}-\gamma \nabla^{a}\left(A^{b}A^{d}\right)\xi_{d}-\frac{1}{2}\gamma A^{d}A^{a}\nabla^{b}\xi_{d}\biggl].
    \end{aligned}
\end{equation}

The other ingredient we need is $\boldsymbol{\xi}\cdot\boldsymbol\Theta$, where the 3-form $\boldsymbol\Theta$ is given by expression~\eqref{ThetaFull}. The components of $\boldsymbol{\xi}\cdot\boldsymbol\Theta$ are obtained as
\begin{equation}
(\boldsymbol{\xi}\cdot\boldsymbol\Theta)_{ij}=\epsilon_{ijab}\left[2 \frac{\partial L}{\partial R_{c e a d}} \nabla_{e} \delta g_{c d}-2 \nabla_{d} \frac{\partial L}{\partial R_{c a d e}} \delta g_{c e}+\frac{\partial L}{\partial(\nabla_{a}A_{d})}\delta A_{d}\right]\xi^{b}.
\end{equation}
Now that we have $\mathbf Q$ and $\boldsymbol{\xi}\cdot\boldsymbol\Theta$ , we are ready to construct the variation $\delta \mathbf Q -\boldsymbol\xi\cdot\boldsymbol\Theta$ required by Wald's formalism in order to obtain the 
entropy of the Schwarzschild black hole solution with a non-trivial vector field, given in Eqs.~(\ref{g1f},\ref{g1a0},\ref{g1pi}) and valid for $\gamma = 1/4$. In this setup, and with a timelike Killing vector field, we find the non-vanishing components of $\mathbf Q$ and $\boldsymbol{\xi}\cdot\mathbf\Theta$ as
\begin{align}
 (\mathbf{Q})_{23}= & \,\frac{\sqrt{h} r^2 \left(f'-16 \pi  A_0 A_0'\right)}{16 \pi  \sqrt{f}}-\gamma\frac{  \sqrt{h} r \left[A_1^2 f h \left(r f'+2 f\right)+A_0 r \left(4 f A_0'-A_0 f'\right)\right]}{f^{3/2}},  \label{q2} \\
(\boldsymbol{\xi}\cdot\mathbf\Theta)_{23}= &\,\frac{1}{2}r\sqrt{\frac{1}{hf^3}}\biggl[{\frac{1}{16\pi}} f\delta h(4f+rf^{'})+{\frac{1}{16\pi}}rh(2f\delta f^{'}-f^{'}\delta f)+\frac{2}{r}hf\delta A_0  \nonumber \\
      &+2\gamma \delta h f\left(P+\frac{Q}{r}\right)^{2}-\frac{2\gamma}{(r-2m)^2}(Q^2+2mrP^2+2rPQ)h^2f\delta f\biggl].\label{ins}%
\end{align}
Taking the variation of $\mathbf Q$ and combining with Eq.~(\ref{ins}), we find
\begin{align}
\int (\delta \mathbf Q - \boldsymbol{\xi}\cdot\boldsymbol\Theta)=& -\frac{1}{2}r\delta h - \frac{2\pi r^{2}A_0\left [ r (\delta f-\delta h)A_{0}'+2(r-2m)\delta A_{0}'\right ]}{r-2m}\nonumber\\
& -2\pi\gamma\Bigg\{ 
\frac{2r^{2}A_{0}}{(r-2m)^{2}}\bigg[ 3m(\delta f-\delta h)+r\delta h-(r-2m)r \delta f' \bigg]\nonumber\\
&+\frac{2}{r}A_{1}\bigg[ 4(r^{2}+2m^{2}-3mr)\delta A_{1}+rA_{1}\Big(3r\delta h-m(\delta f+3\delta h)-r(r-2m)\delta f' \Big) \bigg]\nonumber\\
&-\frac{4rA_{0}}{r-2m}\bigg[2m \delta A_{0}+r^{2}(\delta f-\delta h)A_{0}'+2r(r-2m)\delta A_{0}'\bigg]
+8r^{2}A_{0}'\delta A_{0}
\Bigg\},
\end{align}
where we have evaluated the metric
components $f$ and $h$ of the
Schwarzschild solution. 

In order to evaluate the previous result at asymptotic infinity, we have
\begin{equation}
    \begin{aligned}
        &\delta h(r_{\infty})=\delta f(r_{\infty})=-\frac{2\delta m}{r},\\
        &\delta A_0(r_{\infty})=\delta P+\frac{\delta Q}{r},\\
        & \delta A_{1}(r_{\infty})=\frac{(2mP+Q)(r-2m)r\delta P+(rP+Q)(r-2m)\delta Q+\left[2Q^2+4rPQ+rP^2(r+2m)\right ]  \delta m}{(r-2m)^2\sqrt{Q^2+2rP(mP+Q)}},
    \end{aligned}
\end{equation}
resulting in
\begin{equation}\label{deltahinf}
    \int_\infty (\delta \mathbf Q - \boldsymbol{\xi}\cdot\boldsymbol\Theta)= 
    {\delta m}-4 \pi  {\delta P} m P+4 \pi  {\delta Q} P.
\end{equation}
Now, for evaluating the variation at $r=r_{h}$, we have
\begin{equation}
    \begin{aligned}
        &\delta h(r_{h})=\delta f(r_{h})=-\frac{2m\delta r_{h}}{r_{h}^{2}},\\
        &\delta A_0(r_{h})=\frac{Q\delta r_{h}}{r_{h}^{2}},\\
        & \delta A_{1}(r_{h})=\frac{\left[2m^2P^2+2mPQ+Q^2+P(mP+Q)r_{h}\right]\delta r_{h}}{(r_{h}-2m)^2\sqrt{Q^2+2r_{h} P(mP+Q)}},
    \end{aligned}
\end{equation}
and therefore 
\begin{align}\label{eq:dhp}
  \int_\Sigma (\delta \mathbf Q - \boldsymbol{\xi}\cdot\boldsymbol\Theta)= & \frac{m\delta r_{h}}{r_{h}}
  +\frac{2 \pi  {\delta r}_h \left[P r_h (m P+6 {Q})+8 {Q}^2\right]}{r_h^2}.
\end{align}
Analyzing Eqs.~(\ref{deltahinf}) and~(\ref{eq:dhp}) under the condition $r_h=2m$ imposed by the black hole metric, we find that if $P=2 (3 + \sqrt{13}) Q/r_h$ and $\delta Q=0$, then

\begin{equation}\label{deltahhor}
    \int_\infty (\delta \mathbf Q - \boldsymbol{\xi}\cdot\boldsymbol\Theta) =  \int_\Sigma (\delta \mathbf Q - \boldsymbol{\xi}\cdot\boldsymbol\Theta), 
\end{equation}
 i.e., the first law of black hole thermodynamics proposed by Wald is satisfied. The conditions on the vector charges $P$ and $Q$ are interpreted in the following way: the relation between $P, Q$ and $r_h$ tells us that not all three charges are independent, something that had been pointed out before for vector galileons~\cite{Fan:2017bka,Li:2020kcw}. The condition $\delta Q=0$ tells us that the non-constant part of the vector field component $A_0$ is screened from spacetime not only at the background level (i.e., $Q$ does not appear in the metric) but also at the level of perturbations needed to obtain the first law of black hole thermodynamics. 

Now we can construct the entropy of the Schwarzschild black hole in the vector-tensor theory~(\ref{th}). 
By definition,
\begin{equation}
\int_\Sigma (\delta \mathbf Q - \boldsymbol{\xi}\cdot\boldsymbol\Theta) = \frac{\kappa}{2\pi}\delta S = T\delta S,
\end{equation}
where $\kappa$ is the surface gravity.
Since the black hole is Schwarzschild, we have 
$T=(4\pi r_{h})^{-1}$, and using Eq.~(\ref{deltahhor}) we get
\begin{equation}
    \delta S= 4\pi r_{h}\left[\frac{\delta r_h}{2}+\frac{16 \pi  \left(3 \sqrt{13}+11\right) Q^2 \delta r_h}{r_h^2}\right].
\end{equation}
Noticing that
\begin{equation}
2 r_{h}\delta r_{h}=\delta r_{h}^{2},\quad  \frac{\delta r_{h}}{r_{h}}=\frac{1}{2}\delta \ln r_{h}^{2},
\end{equation}
the change in entropy can 
 be rewritten as
\begin{equation}
    \delta S=\delta\left[ \pi r_{h}^2+
    32 \left(3 \sqrt{13}+11\right) \pi ^2 {Q}^2 \ln \left(\pi  {r_h}^2\right)
    \right].
\end{equation}
Thus we identify the entropy
\begin{equation}\label{eq:entr}
    S= \frac{A}{4}+\frac{1}{4} \pi \tilde\varphi^{2}\ln\frac{A}{4},
\end{equation}
where $A=4\pi r^{2}_{h}$ is the horizon area and we defined $\tilde\varphi = 8 \sqrt{2 \left(3 \sqrt{13}+11\right) \pi } {Q}$. 
Therefore, we conclude that 
the vector-tensor theory~\eqref{th}
introduces a logarithmic term
to the entropy of a Schwarzschild black hole. This type of terms
have been extensively studied in
the literature. A crucial aspect of Eq.~(\ref{eq:entr}) is the sign of the logarithm. If $\tilde\varphi$ is real (equivalently, $Q$ is real), we get the following properties:
\begin{itemize}
    \item For $A/4 > 1$ the entropy of a VG Schwarzschild black hole is larger than that of the
same black hole in GR.
\item For $A/4>1$ the entropy is always positive, but for $A/4<1$, requiring the entropy to be positive implies
\begin{equation}\tilde \varphi^2 < - \frac{A}{\pi \ln \left({A}/{4}\right)}.\end{equation}
\end{itemize}

On the other hand, assuming an imaginary $\tilde\varphi = i \varphi$, with $\varphi\in\mathbb R$ (equivalently, an imaginary $Q$), leads to 
\begin{equation}\label{eq:entrimq}
    S= \frac{A}{4}-\frac{1}{4} \pi \varphi^{2}\ln\frac{A}{4},
\end{equation}
with the following consequences:
\begin{itemize}
\item For $A/4>1$ the entropy of a VG Schwarzschild black hole is smaller than that of the same black hole in GR.
\item Requiring the entropy to be an increasing function of the area leads to the condition $\varphi^2<A/\pi=16 M^2$. This is determined by studying the first derivative of the modified 
entropy, Eq.~(\ref{eq:entrimq}).
\item Requiring the entropy to be always positive leads to $\varphi\leq 2\sqrt{{e}/\pi}$, where $e$ is the Euler 
number. When $\varphi=2\sqrt{{e}/\pi}$ then $S\geq 0$, 
with $S=0$ at $A= 4{e}$.
\end{itemize}
\begin{figure}
\includegraphics[width=0.7\textwidth]{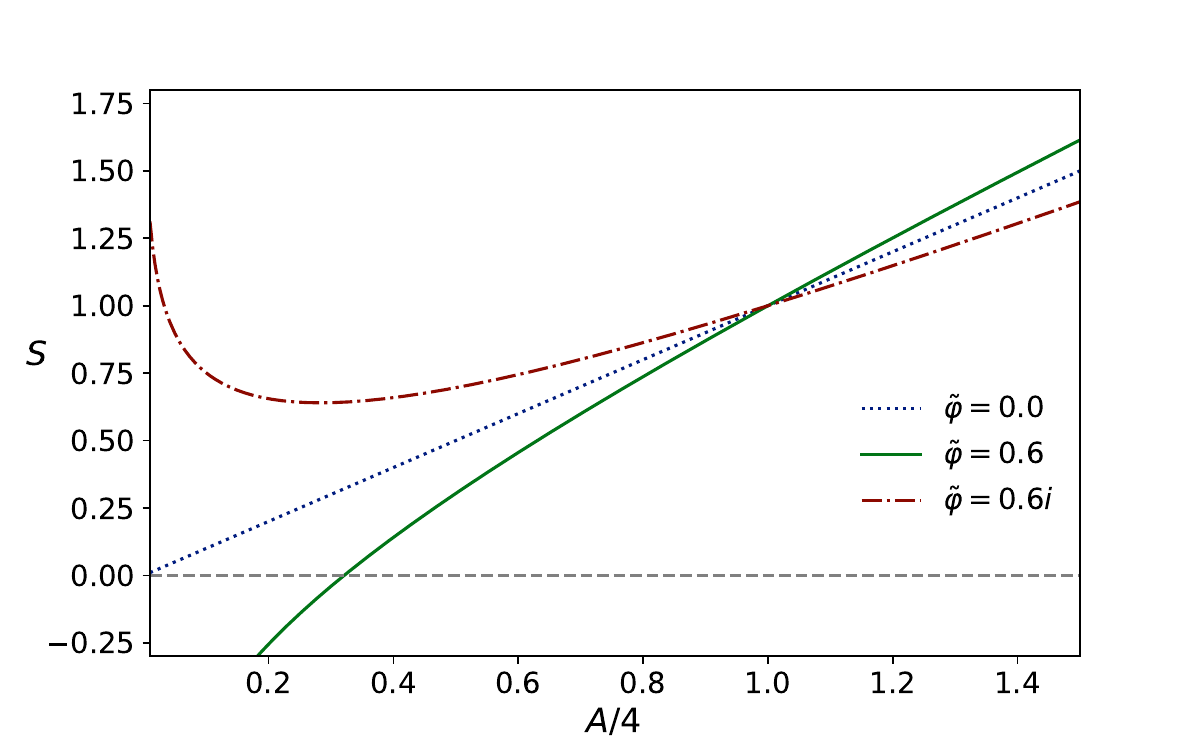}
\caption{Entropy curves for Bekenstein-Hawking ($\tilde\varphi=\varphi=0$), and vector Galileon with real and imaginary $\varphi$. For imaginary $\tilde\varphi$ (imaginary $Q$) there is a region where the entropy decreases as the area increases, while for real $\tilde\varphi$ (real $Q$) there is region with negative entropy.}\label{fig:entropy}
\end{figure}

In Fig.~\ref{fig:entropy} we illustrate the properties mentioned above. 

The possibility of considering real and imaginary $\varphi$ (or equivalently $Q$) arises from the following observation: taking the equations of motion of the
vector galileons for the spherically symmetric ansatz,
Eqs.~(\ref{eqs}), together with the vector field solutions ~(\ref{g1a0},\ref{g1pi}) and with the condition $P \propto Q$ imposed when obtaining the entropy,  we see that if $Q$ is allowed to be imaginary, then $P$ also becomes imaginary, and as a consequence $A_0(r)$ and $\pi(r)$ are pure imaginary numbers, which implies that the energy-momentum tensor of the vector field remains real, since it depends quadratically on $A_0(r), \pi(r)$. Therefore, it
is consistent to consider an imaginary $Q$. 
In the next section we explore other consequences of
the VG modified entropy, arising in the context
of entropic gravity.


\section{Phenomenology}\label{sec:entro}
In this section we 
explore two possible
phenomenological signatures of 
the black hole entropy derived above.
In the first part, we study implications for the
area law, which has been
supported observationally in~\cite{PhysRevLett.127.011103}.
In the second part, we explore consequences
that arise in the context of entropic
gravity. 
\subsection{Area law}
\begin{figure}
\includegraphics[height=0.29\textwidth]{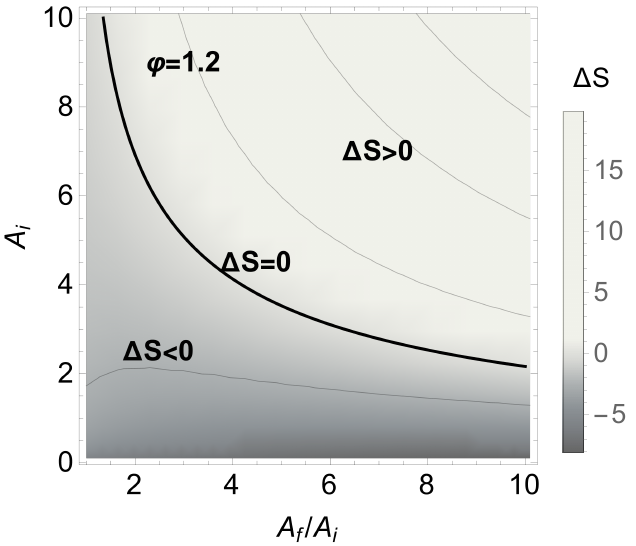}
\includegraphics[height=0.275\textwidth]{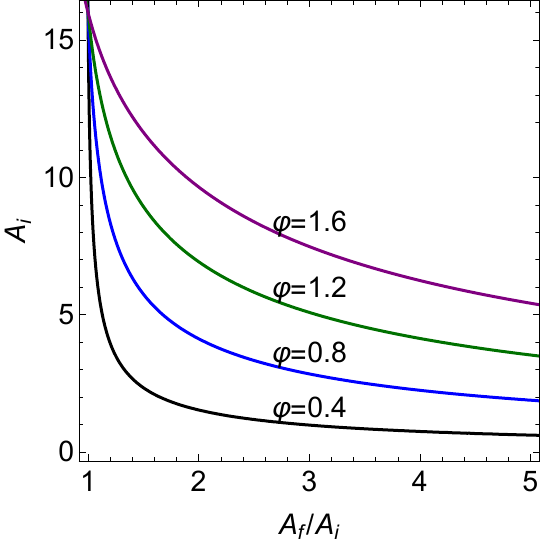}
\includegraphics[height=0.29\textwidth]{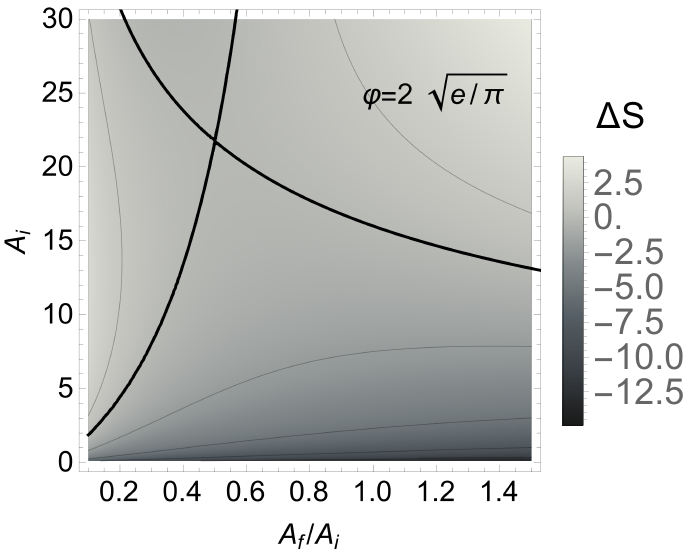}
\caption{Left: regions where an increment of area leads to a negative ($\Delta S<0$) or positive ($\Delta S>0$) change in entropy. The solid line is the curve $\Delta S = 0$. Middle: Curves $\Delta S=0$ for different values of the charge $\varphi$. Right: Processes where $A_f<A_i$ are possible even with a positive change in entropy, which corresponds to the left and top right regions of the plot. }
\label{fig:arealaw}
\end{figure}

In 1971~\cite{Hawking:1971vc} Hawking derived an important theorem {in GR} that shows
that the surface area of a classical black hole can never decrease. This applies also if two black holes merge into a single one: the surface area of the final black hole must be strictly greater than the sum of the areas of the original black holes. It is important to highlight the the area theorem is not in general valid in modified gravity~\cite{Sarkar:2017qln}. Hawking's area theorem is also known as the second law of black hole mechanics, and establishes an
analogy to the second law of thermodynamics and the entropy. The
idea of
testing this theorem with observations
of gravitational waves has been discussed  
in Refs.~\cite{Hughes:2004vw, Giudice:2016zpa, 
PhysRevD.97.124069}, and recently in Ref.~\cite{PhysRevLett.127.011103} it was applied to actual LIGO-Virgo data from GW150914~\cite{LIGOScientific:2016aoc}.  
The recent possibilities for observational
verification of the area theorem makes it
relevant to also investigate theoretical
aspects in the relation between area
and entropy. While for the Bekenstein-Hawking entropy, $S_{BH} = \frac{A}{4}$,
it is true that an increment
of the area leads to an increment of the 
entropy, this is not always the case 
for a modified
entropy, such as Eq.~(\ref{eq:entr}). For instance,
consider two black holes of the same area $A_1$, that
merge into a black hole with a certain final area $A_f$.
Since the charge $Q$ is not a property of the black
hole, we assume that the initial and final entropies are modified by the same factor of $Q$. 
Let us focus on the entropy for imaginary $Q$, Eq.~(\ref{eq:entrimq}), which seems to be more in tension with the area theorem, as it allows for a reduction in area alongside a positive change in entropy.
Thus, for Eq.~(\ref{eq:entrimq}),
the total initial entropy is
\begin{align}
S_i & =\frac{{A_1}}{4}+\frac{{A_1}}{4} -\frac{1}{4} \pi  {\varphi}^2 \ln \left(\frac{{A_1}}{4}\right)-\frac{1}{4} \pi  {\varphi}^2 \ln \left(\frac{{A_1}}{4}\right) \nonumber \\
& = \frac{A_i}{4}-\frac{1}{2} \pi  \varphi^2 \ln \left(\frac{A_i}{8}\right),
\end{align}
and the final entropy is
\begin{align}
    S_f =  \frac{A_f}{4}-\frac{1}{4} \pi \varphi^{2}\ln\frac{A_f}{4},
\end{align}
where $A_i = 2 A_1$. 
Defining $A_f/A_i = \sigma$ ($\sigma>1$ when the area theorem holds) the change
in entropy is
\begin{align}
  \Delta S \equiv   S_f - S_i = \frac{1}{4} \left((\sigma -1) A_i-\pi  \varphi^2 \ln \left(\frac{16 \sigma }{A_i}\right)\right).
  \label{eq:deltas}
\end{align}
Assuming $\sigma>1$ the first term is always positive. However, 
the second term can push $\Delta S$ to negative
values, i.e., the fact that the area grows is not
enough to guarantee that the change in entropy is 
positive. In the left panel of Fig.~\ref{fig:arealaw} we
represent $\Delta S$ 
as a function of the total
area of the initial configuration, $A_i$,
and the ratio $A_f/A_i=\sigma>1$. The charge is fixed at $\varphi=1.2$.  
The black
solid line is the
level curve $\Delta S = 0$, below this line the change in 
entropy is negative. The middle panel of Fig.~\ref{fig:arealaw}
shows the level curves 
$\Delta S = 0$ for different
values of $\varphi$. For each $\varphi$, 
the change in entropy
is similar to the one displayed in the left panel. An important observation,
either from Eq.~(\ref{eq:deltas}) or from Fig.~\ref{fig:arealaw}, is
that for any initial area $A_i>16$ and $\varphi< 2\sqrt{{e}/\pi}$, it is true that
an increase in area leads to
an increase in entropy. For 
smaller $A_i$ the situation
is different, for instance, if
$A_i\sim 5$, a positive
increment of the entropy 
requires that the final area is
around $A_f\sim 1.2 A_i$. Thus,
the area theorem 
$A_f>A_i$ is not enough to
guarantee a never decreasing black hole entropy, instead,
a stronger bound
$A_f> m(\varphi,A_i) A_i$, with $m(\varphi,A_i)>1$ a function
of $\varphi$ and $A_i$ and $m(\varphi, A_i=16) = 1$, is required. The right panel
of Fig.~\ref{fig:arealaw} shows that a positive change in entropy is also possible when the final area is smaller than the initial area.
Finally, it is important to highlight that these details in the relation between area and entropy would be observable only for Planck scale black holes, as can be determined by translating $A_i>16$ to physical units. 

\subsection{Entropic Gravity}
The notion of gravity as an emergent phenomenon was put forward by Jacobson~\cite{Jacobson:1995ab} by noting that the Einstein field equations can be viewed as a thermodynamic equation of state when considering that the  Clausius
 relation $\delta E=TdS$ between heat, temperature and entropy holds for every local Rindler horizon 
(heat is usually denoted by $Q$ in the literature; we use $E$ to avoid confusion with the vector charge $Q$). 
In the spirit of Jacobson's work and in analogy to emergent entropic forces in polymers~\cite{sokolov2010statistical}, Verlinde~\cite{Verlinde:2010hp} proposed that gravity is an effective force which emerges from the entropy via $F\Delta x=T \Delta S$. 
Building on these ideas and motivated by aspects of Loop Quantum Gravity, Modesto and Randono~\cite{Modesto:2010rm} modified Verlinde's assumptions, in particular on the construction of $\Delta S$, arriving at the
following conclusion:
given a test mass $m$ at a distance $R$ of a source of mass $M$, when a change on the entropy of a surface $\mathcal{S}$ of radio $R$ occurs, there appears a force due to that change, given by
\begin{equation}\label{force} 
 \mathbf{F}=-\left.\frac{GMm}{R^{2}}4\ell_{P}^{2}\frac{\partial S}{\partial A}\right|_{A=4\pi R^2} \mathbf{\hat{R}},
\end{equation}
where $\ell_P$ is Planck length.
If one assumes the horizon entropy to be given by the Bekenstein-Hawking formula\footnote{In this section we restore Planck units for the entropy in order to simplify the comparison with existing literature on this topic. For the same reason, in the cosmological analysis we use the relation $\ell_{P}^2 = \hbar G/c^3$.}, $S_{BH}=A/4\ell_{P}^{2}$,
 then Eq.~\eqref{force} yields the Newtonian force. However, it is of interest to consider different entropies. One of the most studied modifications is the well-known logarithmic correction, which has been suggested to be a universal correction term for the Bekeinstein-Hawking entropy,
\begin{equation}
    S(A) = \frac{A}{4\ell_{P}^{2}}-\alpha \ln \frac{A}{4\ell_{P}^{2}}, 
\end{equation}
with $\alpha$ a dimensionless parameter.
For this correction term, the modified Newtonian force and the corresponding potential function have been obtained~\cite{Martinez-Merino:2017xzn}. Comparing with the entropy-area relation~(\ref{eq:entrimq}) for vector Galileons obtained in Sec.~\ref{sec:wald}, we identify
\begin{equation}
\varphi=2\sqrt{\frac{\alpha}{\pi}}.\label{eq:qa}
\end{equation}
The modified force associated to this entropy is
\begin{equation}\label{forcelog}
   \mathbf{F}=-\frac{GMm}{R^{2}}\left[1-\left(\frac{l_p \varphi}{2R}\right)^2\right]\mathbf{\hat{R}}.
\end{equation}
It is worth clarifying the following: evidently, from the
Schwarzschild metric and the equivalence principle, both in GR and vector Galileons
we could deduce that in the weak field limit the trajectory of a particle
is determined by the metric, and therefore by the Newtonian potential $\sim 1/r$
without any corrections. For vector Galileons, this seems incompatible with the corrected
force that arises in the entropic formalism. However, we must emphasise that the correction to
the Newtonian potential derived from the logarithmic entropy is identified as a perturbative  quantum gravity effect~\cite{Modesto:2010rm} (notice that this correction is weighted by the square of the Planck length). Therefore, we do not expect to obtain this correction from a classical weak field analysis. 
The potential function associated to~(\ref{forcelog}) is
\begin{equation}\label{modpot}
    \Phi=-\frac{GM}{R}\left[1-  \frac{1}{3}\left(\frac{\ell_{P}\varphi}{2R}\right)^{2}       \right].
\end{equation}
\begin{figure}[t]
\begin{center}
\includegraphics[width=0.7\textwidth]{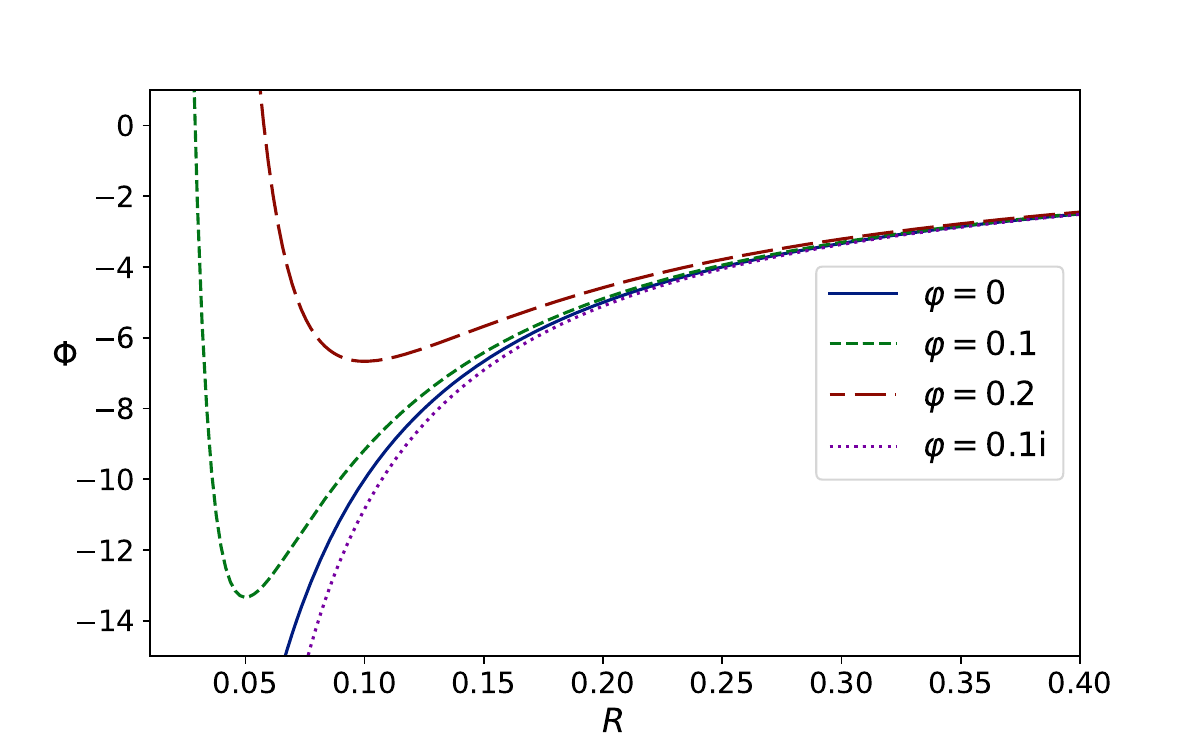}
\caption{Plot of the potential function Eq. \eqref{modpot} for different values of the parameter $\varphi$. When $\varphi\in\mathbb{R}$ and $\varphi\neq 0$ (imaginary $Q$),  the potential develops a minimum, while for imaginary values (real $Q$) it behaves qualitatively like the ordinary Newtonian potential.} \label{pot}
\end{center}
\end{figure}
As discussed before, the theory admits both real and imaginary $Q$ (and therefore $\varphi$), and this determines the relative sign between the terms of the modified potential~\eqref{modpot}. 
A plot of the modified potential for different values of $\varphi$ is shown in Fig.~\ref{pot}. 
Quantitative effects of these (and other) modifications have been studied in the context of planetary orbits~\cite{Perez-Cuellar:2021otp}. Comparing to their results and using Eq.~(\ref{eq:qa}) we see, for instance,
that the correction to Mercury's precession weighted by $\varphi^2$ is around $10^{84}$ times smaller than
the observational result. Thus, extremely large vector charges would be needed to get a  relevant contribution
from the logarithmic correction. For modified forces and potentials obtained from an entropy-area relation with further correction terms, see~\cite{Diaz-Saldana:2018ywm}. 

 The ideas of Jacobson have also been applied in cosmological scenarios~\cite{Cai:2008ys,Diaz-Saldana:2018gxx,Sheykhi:2010zz}. In~\cite{Cai:2008ys}, this is done by considering the Clausius relation on the apparent horizon of a Friedmann-Robertson-Walker (FRW) universe and assigning a given temperature and entropy to the horizon. The energy flow on the apparent horizon is determined by the matter content, which is considered to be a perfect fluid. 
 If one assumes the horizon entropy to be given by the Bekenstein-Hawking formula, then the Clausius relation together with the continuity equation yield the Friedmann equation for the FRW universe. On the other hand, if one considers a modified entropy-area relationship on the horizon, then a modified Friedmann equation {is} obtained.  
 In~\cite{Cai:2008ys} the following modified Friedmann equation for a flat universe arising in the context of quantum cosmology is considered,
 \begin{equation}\label{mf}
    H^2=\frac{8 \pi G}{3} \rho\left(1-\frac{\rho}{\rho_{\mathrm{crit}}}\right).
\end{equation}
Due to the modification term in this equation, the big bang singularity is replaced by a quantum bounce happening at $\rho=\rho_{\mathrm{crit}}$. The modified Friedmann Eq.~\eqref{mf} can be obtained from the modified entropy-area relation
\begin{equation}\label{entrcai}
    S=\frac{A}{4 \ell_{P}^{2}}+\frac{3}{4 \pi G^2 \rho_{\text {crit }}} {\frac{c^5}{\hbar}}\ln \frac{A}{4 \ell_{P}^{2}},
\end{equation}
where $A$ is the area of the horizon. Although this form of the entropy is the familiar one with the logarithmic correction, in this case the correction term gives an opposite  contribution to the entropy, since the prefactor in the logarithmic term is positive. We see that this entropy is equivalent to~\eqref{eq:entr} with the identification of a real $\varphi$,
\begin{equation}
   \tilde{\varphi}^2=\frac{4}{\pi}\left(\frac{3}{4 \pi G^2 \rho_{\text {crit }}}{\frac{c^5}{\hbar}}\right).
\end{equation}
\begin{figure}[t]
\begin{center}
\includegraphics[width=0.6\textwidth]{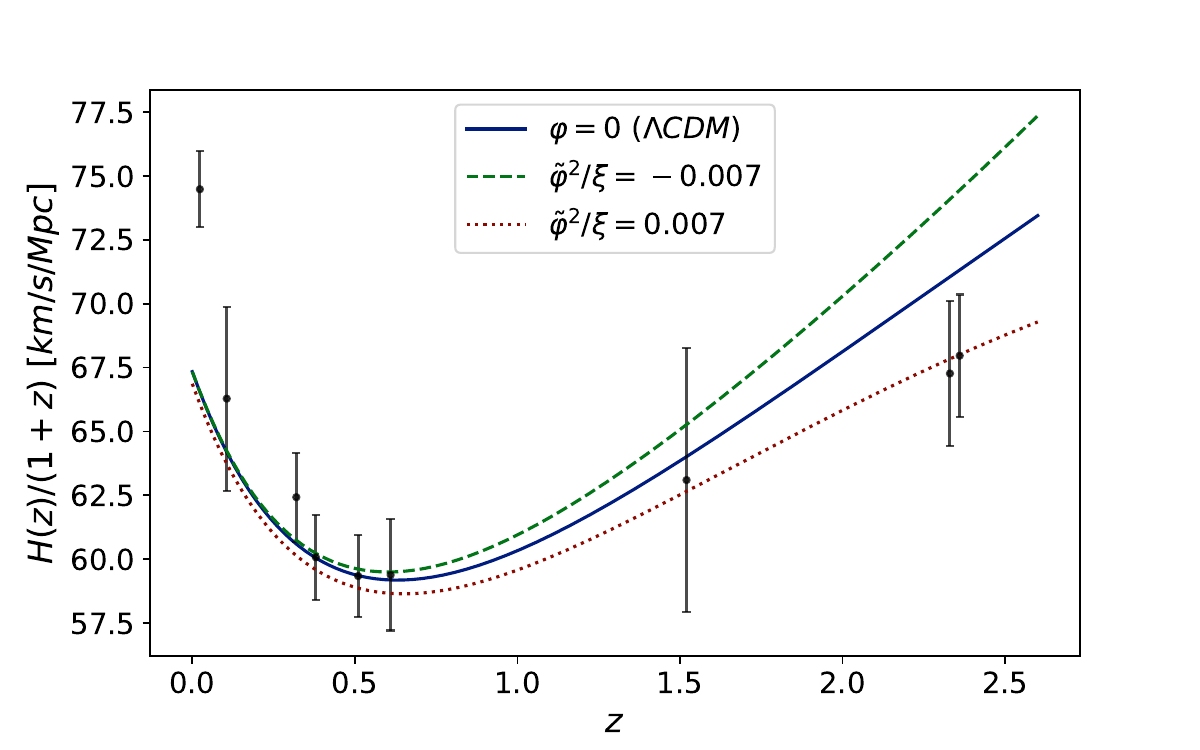}
\caption{Plot of $H(z)/(1+z)$ for the modified Friedmann Eq.~\eqref{modh} that comes from the entropy (\ref{eq:entr}) (dotted). We also plot the case that corresponds to the entropy (\ref{eq:entrimq}) (dashed).  Points and errors bars are taken from observational Hubble data compiled in~\cite{Yu:2017iju}.}\label{fig:hz}
\end{center}
\end{figure}
In order to study some consequences of the modified Friedmann Eq.~\eqref{mf}, let us express the density as $\rho=\rho_{M}+\rho_{\Lambda}$, where $\rho_{M}$ stands for non relativistic matter $\rho_{M}(z)=\rho_{0M}(1+z)^{3}$, and $\rho_{\Lambda}$ for the constant dark energy density $\rho_{\Lambda}=\Lambda/8\pi G$, and let us introduce the density parameters {at present time,}
\begin{equation}
    \Omega_{0M}=\frac{8\pi G \rho_{0M}}{3 H_{0}^{2}},\quad \Omega_{0\Lambda}=\frac{\Lambda}{3 H_{0}^{2} } .
\end{equation}
With this, we can rewrite Eq.~\eqref{mf} as 
\begin{align}\label{modh}
   \frac{H(z)^2}{H_0^2} = & \Omega_{0M}(1+z)^{3}+\Omega_{0\Lambda}\left(1-\frac{\tilde{\varphi}^{2}}{\xi}\Omega_{0\Lambda}\right)-2\frac{\tilde{\varphi}^{2}}{\xi}\Omega_{0M}\left[\Omega_{0\Lambda}+\frac{\Omega_{0M}}{2}(1+z)^{3}\right](1+z)^{3},
\end{align}
where we have defined the dimensionless parameter
\begin{equation}
\xi=\frac{8}{\pi G H_{0}^{2}}{\frac{c^5}{\hbar}}.
\end{equation}
Note that in the limit $\tilde{\varphi}\to 0$ the usual Friedmann equation is recovered. In figure~\ref{fig:hz} we show a plot of $H(z)/(1+z)$ for different values of $\tilde{\varphi}^{2}/\xi$.
We see that the modifications to the Hubble factor are larger for large redshift, and
depending on whether $Q$ is real ($\tilde\varphi\in\mathbb R$) or imaginary ($\varphi\in\mathbb R$), the values of $H(z)$ are 
 smaller or larger, respectively, than in $\Lambda$CDM. The values of $\tilde{\varphi}$ used for
this figure are exaggerated in order to make their effect noticeable. Since $\xi\sim10^{122}$, extremely large values for $\tilde{\varphi}$ would be needed in order to obtain relevant modifications to the evolution of the Hubble factor.

\section{Conclusions}\label{sec:dis}
We studied the entropy of a Schwarzschild black hole dressed with a non-trivial vector field in a model belonging to generalised Proca theories, also known as vector Galileons. The non-trivial vector field has both a temporal and a radial component, whose profiles depend on two integration constants. In principle, these integration
constants are independent; however, consistency with the first law of black hole mechanics demands a relation
between these constants and the black hole mass, leaving only two independent quantities that can be identified as charges of the black hole and vector field. The
fact that some conditions need to be imposed on  
the parameters of the solutions of vector Galileon theories had been pointed out in~\cite{Fan:2017bka,Li:2020kcw}, but a 
complete calculation had not been performed so far. 

After imposing the appropriate constraints, we find that the entropy matches the Bekenstein-Hawking relation with a
logarithmic contribution modulated by the independent charge of the vector field. {This is relevant because logarithmic corrections to the entropy are considered to be universal. }
A logarithmic term in the entropy is usually associated to quantum corrections, 
whose back-reaction on the black hole geometry needs to be taken into account in order to get a fully consistent analysis. However, in our results the logarithmic correction
arises from a classical action and an exactly Schwarzschild background, signalling 
the existence of particular cases  where additional fields of the theory protect the metric from the back-reaction expected 
from logarithmic corrections. It would be interesting to
explore this idea in the context of effective field theories that have been recently
applied to obtain the back-reaction on the metric~\cite{Xiao:2021zly,Calmet:2021lny}.

We analysed some consequences of the logarithmic contribution to the entropy. Since our
calculation is classical, in principle  the logarithmic part of the entropy does not need to be treated as a small correction term. Demanding that the entropy is always increasing, we find
an upper bound on the vector charge, of the form $-Q^2\lesssim{A}$ (in the case where $Q$ is pure imaginary). This still allows for the possibility of large vector charges in large black holes, raising the question
of whether this modified entropy could be verified experimentally. An interesting direction along these lines is the observational corroboration of the area theorem of GR, which exploits the fact that a consequence of this theorem is that when two black holes merge into a single one, the area of the final black hole is larger than the sum of the areas of the initial configurations. In modified gravity, the area theorem is not in general valid. Assuming that the entropy is the quantity that needs to be always increasing, we find that the condition on the area can be relaxed. i.e., a decreasing area can be compatible with an increasing modified entropy. Thus, an observation of a decreasing area would be problematic for GR, but not for theories with logarithmic corrections to the entropy. However, our results indicate that this is only an issue for Planck scale black holes. At these scales, we expect quantum corrections to modify our predictions.  

In addition, following the arguments of entropic gravity, the
logarithmic correction to the entropy translates into
a modified Newtonian potential, which has consequences, e.g. for planetary orbits~\cite{Perez-Cuellar:2021otp}. 
Finally, also in the context of entropic gravity, we review the modified Friedmann equations
resulting from the logarithmic correction to the entropy, and we use there results to analyse cosmological effects of the vector charge.  Both in the astrophysical and cosmological scenarios, we find that the corrections due to the vector charge are several orders of magnitude smaller than the classical scales of the respective problems.

\section*{Acknowledgements}
{\bf J.C.L.D and J.C.} are supported by CONAHCyT/DCF-320821.  {\bf I. D. S} is supported by  CONAHCyT/Estancias Posdoctorales por M\'exico. {\bf C.M.R.} was supported by CONAHCyT PhD scholarship program.
 It is a pleasure to thank Antonio P\'erez Cortés for his important contribution to the early stages of this project.

\bibliographystyle{unsrt}
\bibliography{refs}

\end{document}